\documentclass[12pt,a4paper,oneside,titlepage]{article}
\usepackage{geometry}
\usepackage[ngerman, english]{babel}
\usepackage[utf8x]{inputenc}
\usepackage{ucs}
\usepackage{amsmath}
\usepackage{amsfonts}
\usepackage{amssymb}
\usepackage{url}
\usepackage{graphicx}
\usepackage{microtype}
\usepackage{lipsum}
\usepackage{booktabs}
\usepackage{caption}
\usepackage{wasysym}
\usepackage[toc,page]{appendix}
\usepackage{braket}
\usepackage{authblk}
\usepackage{subfig}
\usepackage{comment}

\usepackage[normalem]{ulem} 
\usepackage[utf8x]{inputenc} 

\title{Strong coupling between organic dye molecules and lattice modes of a dielectric nanoparticle array}
\author[*]{Rebecca Heilmann}
\author[*]{Aaro I. V{\"a}kev{\"a}inen}
\author[*]{Jani-Petri Martikainen}
\author[*]{P{\"a}ivi T{\"o}rm{\"a}}
\affil[*]{Department of Applied Physics, Aalto University School of Science, P.O. Box 15100, Aalto, FI-00076, Finland}

\begin{document}

\maketitle

\begin{abstract}
Plasmonic structures interacting with light provide electromagnetic resonances which result in a high degree of local field confinement enabling the enhancement of light-matter interaction. Plasmonic structures typically consist of metals which, however, suffer from very high ohmic losses and heating. High-index dielectrics, on the other hand, can serve as an alternative material due to their low-dissipative nature and strong scattering abilities.
We study the optical properties of a system composed of all-dielectric nanoparticle arrays covered with a film of organic dye molecules (IR-792), and compare these dielectric arrays with metallic nanoparticle arrays. We tune the light-matter interaction by changing the concentration in the dye film and report the system to be in the strong coupling regime. We observe a Rabi splitting between the surface lattice resonances (SLRs) of the nanoparticle arrays and the absorption line of the dye molecules of up to 253 meV and 293 meV, for the dielectric and metallic nanoparticles, respectively. The Rabi splitting depends linearly on the square root of the dye molecule concentration , and we further assess how the Rabi splitting depends on the film thickness for a low dye molecule concentration. For thinner films of thicknesses up to 260 nm, we observe no visible Rabi splitting. However, a Rabi splitting evolves at thicknesses from 540 nm to 990 nm. We perform finite-difference time-domain simulations to analyze the near-field enhancements for the dielectric and metallic nanoparticle arrays. The electric fields are enhanced by a factor of 1200 and 400, close to the particles for gold and amorphous silicon, respectively, and the modes extend over half a micron around the particles for both materials.\par
\end{abstract}

\section{Introduction}
 If an emitter and an optical mode of its local environment are coupled to each other and exchange energy, hybrid states of the original modes can arise for a sufficiently strong coupling. A system is in the strong coupling regime when the coupling, or Rabi splitting, exceeds the linewidth of the hybrid modes. \par
 Strong coupling has been observed between single emitters and high quality cavity modes, and in various other systems, for a brief review see the introduction of Ref.~\cite{torma_strong_2015}. In this work, we focus on strong coupling in the context of plasmonics~\cite{torma_strong_2015} where the strong coupling typically involves ensembles of emitters, although the single emitter regime has been approached as well \cite{kravets2018plasmonic}. For ensembles of emitters, the Rabi splitting is in general proportional to the square root of their concentration. In the strong coupling regime of plasmonic systems, coherence phenomena can be observed ~\cite{guebrou2012coherent,shi_spatial_2014}.  Plasmonic systems in the strong coupling regime have been reported to form polariton lasers and condensates ~\cite{vakevainen2019sub}, exhibit enhanced conductivity \cite{orgiu2015conductivity} and efficient second harmonic generation \cite{chervy2016high}. Light strongly coupled to molecular transitions has also been reported to alter chemical reactions ~\cite{ribeiro2018polariton,hertzog2019strong}. \par
An interesting plasmonic system where strong coupling has been observed and utilized are metal nanoparticle arrays. In these arrays, the localized plasmonic resonances of the nanoparticles hybridize with diffracted orders of the periodic structure, forming lattice modes called surface lattice resonances (SLRs) ~\cite{de2007colloquium,bozhevolnyi2017quantum,wang2018rich}. The SLR modes couple to the absorption line of emitters, for example organic dye molecules, and reach the strong coupling regime as evidenced by the square root dependence of the Rabi splitting on the emitter concentration as reported by Väkeväinen \textit{et al.} \cite{vakevainen_plasmonic_2014}. Strong coupling in such arrays has already led to new coherence, lasing and condensation phenomena ~\cite{shi_spatial_2014,vakevainen2019sub,ramezani2017plasmon,de2018interaction}.\par
For the further exploitation of strong coupling in nanoparticle arrays, it would be desirable to increase the lifetime of the system, and eventually make them complementary metal-oxide-semiconductor (CMOS) compatible for integration with present microelectronic components. The metal nanoparticles have a major drawback: high absorption losses which lead to broad resonances and short lifetimes.  High-index dielectric materials on the other hand offer low-dissipative properties and provide not only electric but also magnetic resonances due to the excitation of circular displacement currents inside the material \cite{staude2017metamaterial}.\par
Many plasmonic effects observed in metallic materials can be realized with high-index dielectric materials. Optically induced resonances have been realized with different shapes of dielectric nanoparticles, such as spheres, spheroids, disks, rings and cylinders \cite{kuznetsov_optically_2016}. Dielectric nanoparticle arrays have resonances analogous to the localized surface plasmon resonances and diffracted orders, which form SLRs similar to those observed in metallic nanoparticle arrays \cite{evlyukhin2010optical}.\par
Silicon, in particular, has advantages such as its well established fabrication technologies, CMOS-compability and the possibility to integrate it into existing silicon optics, as well as being low-cost compared to traditional plasmonic materials such as gold or silver \cite{staude2017metamaterial}. In addition, silicon nanoparticles are promising candidates for spectroscopic measurements close to the particles due to their low heat generation \cite{krasnok2018spectroscopy}. Strong coupling in silicon nanostructures has been predicted theoretically between J-aggregates and silicon nanospheres \cite{tserkezis2018mie} and  demonstrated experimentally in similar structures \cite{wang2016resonance}, as well as in systems of J-aggregates and silicon nanodisks \cite{liu2018resonance,todisco2019magnetic}. \par
In this work, we explore the strong coupling phenomena in dielectric nanoparticle arrays. The cylindrical nanoparticles consist of amorphous silicon and are arranged in square lattices. Organic dye molecules (IR-792) act as emitters. Transmission measurements are conducted with the amorphous silicon nanoarrays, and, for comparison, with similar gold nanoarrays. The concentration of the dye molecules is varied and the Rabi splitting extracted. To further investigate differences between amorphous silicon and gold nanostructures we perform FDTD simulations using commercially available software \cite{lumerical}. We find strong coupling between the SLRs of dielectric nanoarrays and the absorption line of the emitters.\par
We observe an increase in the Rabi splitting with the square root of the dye concentration. The amorphous silicon and gold nanoparticle arrays show similar Rabi splittings of 253 meV and 293 meV, respectively. This makes amorphous silicon nanoparticle arrays useful for applications as metal nanoparticle arrays.\par

\section{Materials and methods}
\begin{figure}[htbp!]
\centering
\includegraphics[width=0.8\textwidth]{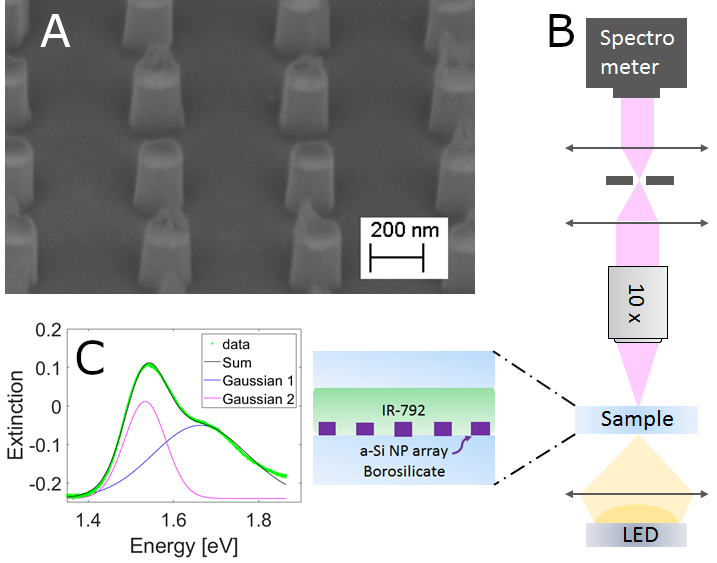}
\caption{(A) SEM picture of an amorphous silicon sample before chromium etching. (B) Transmission measurement setup and schematic of the sample with the dye. (C) Absorption spectrum of IR-792 with a concentration of 100 mM solved in PMMA A3. We fit two Gaussians to the spectrum, shown by the pink and blue curves. The pink curve corresponds to the main absorption peak. }
\label{fig:setup}
\end{figure}

We fabricate arrays of cylindrical gold and amorphous silicon nanoparticles. The gold nanocylinders have a diameter of 100 nm and a height of 50 nm arranged in a square lattice, while the amorphous silicon nanocylinders have a diameter of 100 nm and a height of 100 nm with the same lattice dimensions. In both cases the lattice periodicities are varied in 10 nm steps from 580 to 610 nm in x- and y-direction. Figure \ref{fig:setup} A shows a scanning electron microscope (SEM) picture of an amorphous silicon nanoarray.\par
The samples are fabricated on borosilicate substrates with electron-beam lithography. To fabricate the gold samples, a resist layer of PMMA A4 is spin-coated onto the substrate and 10 nm of aluminium is evaporated on top of the resist to obtain a conductive layer for the electron beam exposure. The layout for the nanoparticle arrays is patterned into the resist. The aluminium layer is etched in AZ 351 B and the resist is developed in a IPA:MIBK (1:3) solution. After this, 2 nm of titanium and 50 nm of gold are evaporated onto the sample and a lift off is carried out in acetone. The titanium layer is necessary to enhance the adhesion between the substrate and gold.\par
For the fabrication of the amorphous silicon samples, 100 nm of amorphous silicon is deposited onto the substrate with chemical vapor deposition. Following this, chromium nanoarrays are fabricated on top of the amorphous silicon similarly to the gold nanoparticles. These chromium nanoarrays serve as masks for the etching of amorphous silicon via reactive-ion-etching. The reactive-ion-etching is conducted at a power of 200 W, a pressure of 10 mTorr, with a SF$_{6}$-flow of 15 sccm and a CHF$_{3}$-flow of 50 sccm for 5 minutes. The chromium mask is etched off afterwards in chromium wet etchant, so that only the amorphous silicon nanoarrays are left on the sample. \par
The IR-792 molecules are dissolved into PMMA A3 (3\% PMMA in anisol) with different concentrations and spin-coated onto the samples at 3000 rpm for one minute, which gives rise to a film thickness of approximately 150 nm. The arrays are embedded in a symmetric refractive index environment by immersion in index matching oil and addition of another borosilicate slide on top (n = 1.52). \par
Different film thicknesses of the dye films are obtained by changing the viscosity of the PMMA mixture and varying the spin-coating speed. The higher the amount of PMMA, the more viscous the mixture becomes. The dye molecules are dissolved into PMMA-anisole mixtures with 3\% and 6\% PMMA in anisole and spin-coated with speeds of 1000, 2000 and 3000 rpm. The film thicknesses are measured with a profilometer.\par
The transmission measurement setup is depicted in Figure \ref{fig:setup} B. The sample is illuminated by a white light source, collected by a 0.3 NA objective and guided to a spectrometer. The light excites in-plane propagating modes in $k_{x}$- and $k_{y}$-direction, each of which have two polarization components. The modes with the E-field polarized perpendicular and parallel to the propagation direction are transverse electric (TE) and transverse magnetic (TM) modes, respectively. With the spectrometer, we only record the modes propagating in $k_{y}$-direction and a linear polarizer is used to filter out the TM modes. Hence, only the TE modes are visible in the measured extinction spectra. \par
To characterize the Rabi splitting between the SLR and the absorption line of the dye, we fit a coupled modes model to the measured extinction spectra. First, a three-coupled-modes model is used to obtain the SLRs of the arrays without the dye. The modes involved are the localized surface plasmon resonances (LSPRs) and the two diffracted orders. Since only the upper branches, i.e. the modes above the diffracted-order crossing, are involved in the coupling, it is sufficient to only take these into consideration. After obtaining the SLRs of the bare arrays, a two-coupled-modes model containing the SLRs and the main absorption peak of the IR-792 is fitted to the measured extinction spectra. The coupled modes model between the SLRs and the main absorption peak of the IR-792 is given by the matrix
\begin{equation}
M =
\begin{pmatrix}
E_{\mathrm{SLR}}-i \gamma_{\mathrm{SLR}}+s & \Omega/2 \\
\Omega/2 & E_{\mathrm{dye}}- i \gamma_{\mathrm{dye}}
\end{pmatrix},
\end{equation}
where $E$ and $\gamma$ are the energies and line widths of the uncoupled modes, respectively, and $\Omega$ is the Rabi splitting. In addition, we introduced a constant shift, $s$, to account for a shift in the refractive index introduced by the dye molecule film. Here, the modes of the coupled system are given by the eigenvalues of the matrix. We extract the peaks of the SLRs in the measured extinction spectra for each k and apply a least-square-sum model to minimize the difference between the eigenvalues of the matrix and the extracted peaks from the measurements. Here, $\Omega$ and $s$ are used as fitting parameters. Figure \ref{fig:setup} C shows the absorption spectrum of IR-792 at a concentration of 100 mM. The absorption spectra at the other concentrations used are shown in the Supplementary Material, Figure 2.\par
The optical properties of the dielectric and gold nanoarrays are studied with the finite-difference time-domain (FDTD) method by Lumerical’s commercially available software. The arrays consist of cylinders in diameter and height as in the experiments. The refractive indices for gold (“Au(Gold) - CRC”) and titanium ("Ti(Titanium) - CRC") are used from the material database provided by Lumerical which are based on the CRC Handbook of Chemistry and Physics \cite{lide1995crc}. The refractive index for amorphous silicon is added as a custom made material according to the measured refractive index of amorphous silicon by Li \textit{et al.} \cite{li_engineering_2018}. The background refractive index is set to 1.52. To simulate the array, periodic boundary conditions are used around the unit cell with perfectly matched layers normal to the array to absorb the excitation and scattered waves. The mesh size is set to 5 nm in all directions over the whole simulation region. The excitation source is a plane wave source with a wavelength range from 550 to 1000 nm with 225 frequency points. The light is polarized orthogonally to the nanocylinder axis.\par

\section{Results}
\begin{figure}[htbp!]
\centering
\includegraphics[width=0.8\textwidth]{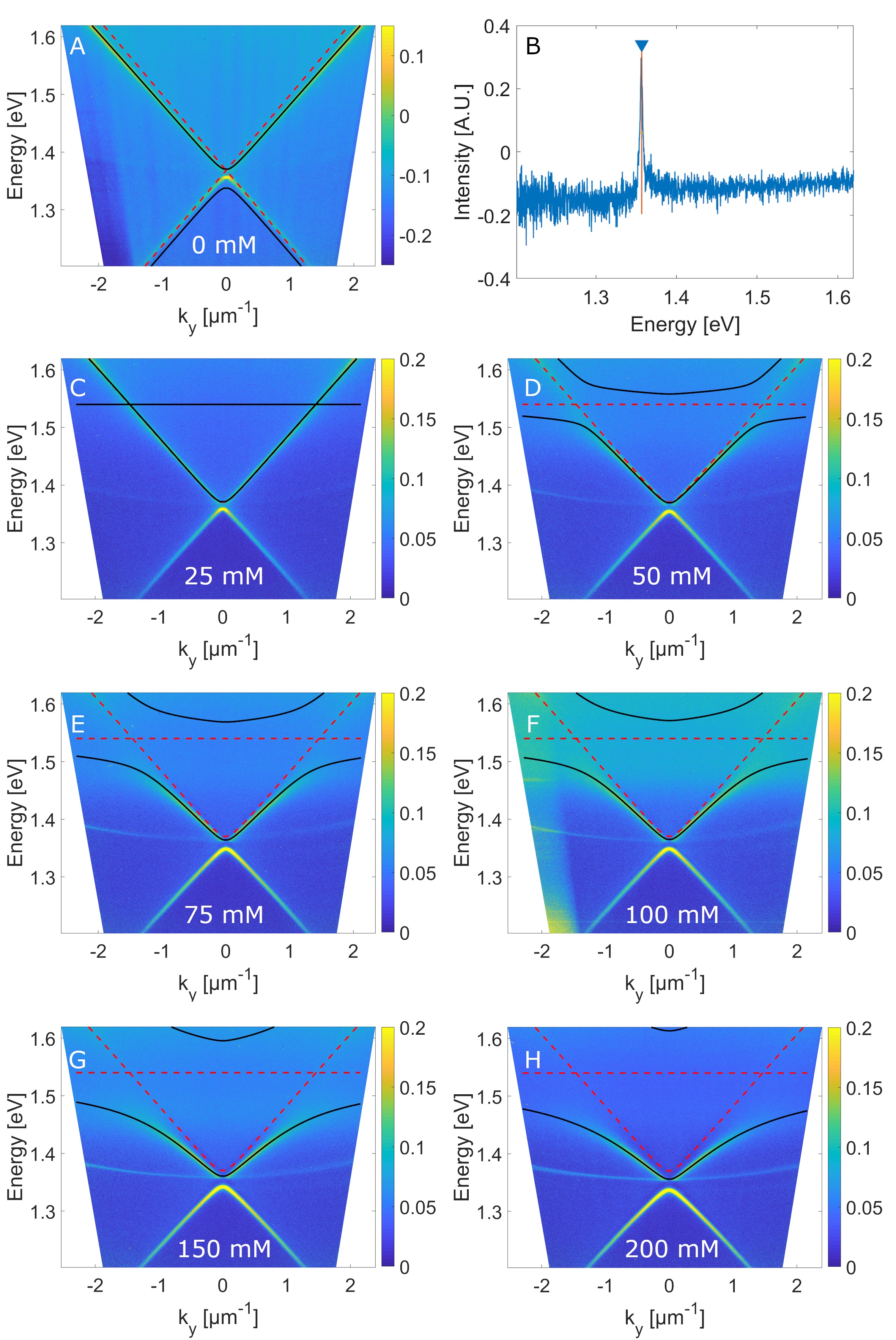}
\caption{Dispersion measurements of a nanoarray of amorphous silicon nanoparticles with a periodicity of 600 nm in x- and y-direction, diameter of 100 nm and dye concentrations of (A) 0 mM (C) 25 mM, (D) 50 mM, (E) 75 mM, (F) 100 mM, (G) 150 mM and (H) 200 mM. The film thickness is 150 nm. Maximum extinction is given in yellow. (B) shows a crosscut through (A) at $k_{y}=0$. In (A) the red dashed lines depict the uncoupled diffracted orders of the lattice and the black solid lines the SLRs obtained by the coupled mode fitting of the extinction maxima. In (C), (D), (E), (F), (G) and (H) the horizontal red dashed line depicts the absorption line of the dye molecules and the diagonal red lines correspond to the fitted SLRs (upper branch) from Figure \ref{fig:splitting} (A). The solid black lines depict the coupled mode fitting of the extinction maxima. It is clearly visible that the Rabi splitting increases with increasing dye molecule concentration.}
\label{fig:splitting}
\end{figure}

Figure \ref{fig:splitting} shows the measured extinction spectra in $k$-space for an amorphous silicon nanoarray without and with dye molecules  at concentrations of 25 mM, 50 mM, 75 mM, 100 mM, 150 mM, and 200 mM. Figure \ref{fig:splitting} A clearly shows the uncoupled SLRs of the array and Figure \ref{fig:splitting} C, although dye molecules at a concentration of 25 mM are present, does not show any difference to the uncoupled modes. As the dye concentration is increased further, see Figures \ref{fig:splitting} D-H, the system develops an anticrossing at the point where the energy of the dye absorption maxima corresponds to the SLR energy. With increasing dye concentration, the anticrossing or Rabi splitting increases and is expected to increase linearly with the square root of the dye concentration \cite{torma_strong_2015,bozhevolnyi2017quantum}. Figure \ref{fig:splitting} B shows a crosscut through Figure \ref{fig:splitting} A at $k_{y}=0$. The peak in the spectrum is the $\Gamma$-point at 1.36 eV with a full width at half maximum of 9 meV. The corresponding measurements taken for the gold nanoparticle arrays are shown in Figure 1 of the Supplementary Material. \par
To characterize the Rabi splitting, a coupled modes model is applied for gold and amorphous silicon nanoarrays as described in the Materials and Methods section. The Rabi splitting is calculated for gold and amorphous silicon arrays with periodicities ranging from 580 nm to 610 nm in 10 nm steps. Sufficiently away from the single particle resonance, the SLR modes for different periodicities have similar density of states and thus couple in the same way with the molecules. Therefore, we can use the standard deviation of the values obtained for different periodicities, at the same concentrations, in estimating the error limits for the Rabi splitting. \par
\begin{figure}[htbp!]
\centering
\includegraphics[width=0.8\textwidth]{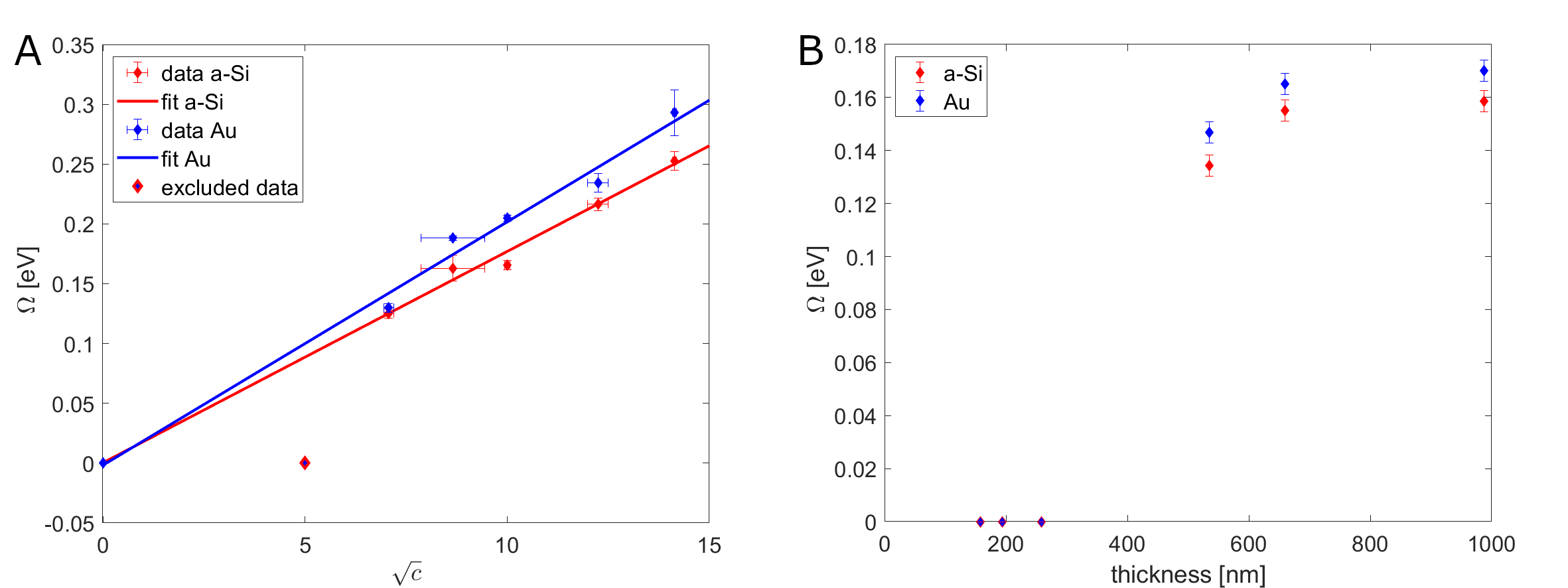}
\caption{Dependence of the Rabi splitting on the square root of the concentration $\sqrt{c}$ for a film thickness of 150 nm (A) and Rabi splitting for $c$=25 mM as a function of different film thicknesses (B). In (A), the data point at $c=25$ mM is excluded since the splitting is smaller than the linewidth and thus not measurable.}
\label{fig:strong_coupling}
\end{figure}

The strong coupling condition in the case of two modes with finite linewidths is given by $\Omega > \frac{\gamma_{\mathrm{dye}}}{2} +\frac{\gamma_{\mathrm{SLR}}}{2}$, where $\Omega$ is the Rabi splitting and $\gamma_{\mathrm{dye}}$ and $\gamma_{\mathrm{SLR}}$ are the linewidths of the uncoupled dye molecules and SLRs, respectively \cite{torma_strong_2015}. It is evident from Figure \ref{fig:splitting} A that the uncoupled SLR width stays constant for all $k$. The value for $\gamma_{\mathrm{SLR}}$ was calculated to be 10 meV for both amorphous silicon and gold as obtained by taking a crosscut at $k=0.5$ $\mu\mathrm{m}^{-1}$. To obtain $\gamma_{\mathrm{dye}}$, the sum of two Gaussians is fitted to the measured absorption spectra of the dye for each concentration. From this fit, the full-width at half maximum of the main absorption peak is extracted. The obtained values for $\Omega$ and $\gamma_{\mathrm{dye}}$ are listed in Table \ref{tab:splitting}. The dye linewidth can be alternatively determined by fitting one Gaussian to the absorption spectra, see corresponding results in the Supplementary Material (Table 1 and Figure 3); while the numerical values of the Rabi splitting and the linewidths are slightly changed, the qualitative behaviour is the same. The strong coupling condition is met for all concentrations except 25 mM. Therefore, and since there is no visible splitting for a concentration of 25 mM (see Figure \ref{fig:splitting} C), it is concluded that the system with such a low concentration is not in the strong coupling regime (splitting below the linewidth). Hence, we exclude this particular data point in Figure \ref{fig:strong_coupling} A.\par
Figure \ref{fig:strong_coupling} A shows the Rabi splitting as a function of the square root of the molecular concentration, $\sqrt{c}$, and a linear dependence is clearly evident. Both systems, gold and emitters as well as amorphous silicon and emitters, are in the strong coupling regime. In general, the Rabi splitting is larger for the gold than for the amorphous silicon nanoparticle arrays.\par
Now, one has to keep in mind that quite thin films were used for the above results. If the film covers the whole mode volume of the plasmonic system, then the concentration determines the Rabi splitting. However, if the mode volume extends beyond the film thickness, then by increasing the thickness, one introduces more emitters into the mode volume and the Rabi splitting may increase with the film thickness also for the same concentration. Therefore, by varying the film thickness, one can obtain indirect information about the extent of the mode volume which is in general a difficult question to address in plasmonics ~\cite{ruppin2002electromagnetic,koenderink2010use}. \par
To investigate this issue, we measure different dye layer thicknesses with 25 mM dye concentration, as shown in Figure \ref{fig:strong_coupling} B. With thicker dye layers, the system evolves into the strong coupling regime. 
It is clear that the splitting increases with the dye layer thickness, suggesting that the amount of emitters participating in the interaction with the SLR increases. The Rabi splitting arises for a 530 nm thick dye layer but its size does not increase much when increasing the layer thickness further. The saturation of the Rabi splitting as a function of dye layer thickness implies that the SLR mode volume, despite having high field enhancement near the nanoparticles, reaches to some extent up to 700 nm from the array plane. It has been predicted theoretically \cite{gonzalez2013theory} and observed in silver hole arrays coupled to CdSe nanoplatelets \cite{winkler2018room} that the Rabi splitting first increases linearly and then saturates with increasing film thickness. The surface plasmon polariton modes decay exponentially from the metal, and additional emitters cannot couple to the modes beyond the confinement length. 
Since the amorphous silicon particles show in general similar results as the gold particles, we conclude that the modes in the dielectric nanoparticle arrays behave similarly to the modes of the metal nanoparticle structures.\par
\begin{figure}[htbp!]
\centering
\includegraphics[width=0.8\textwidth]{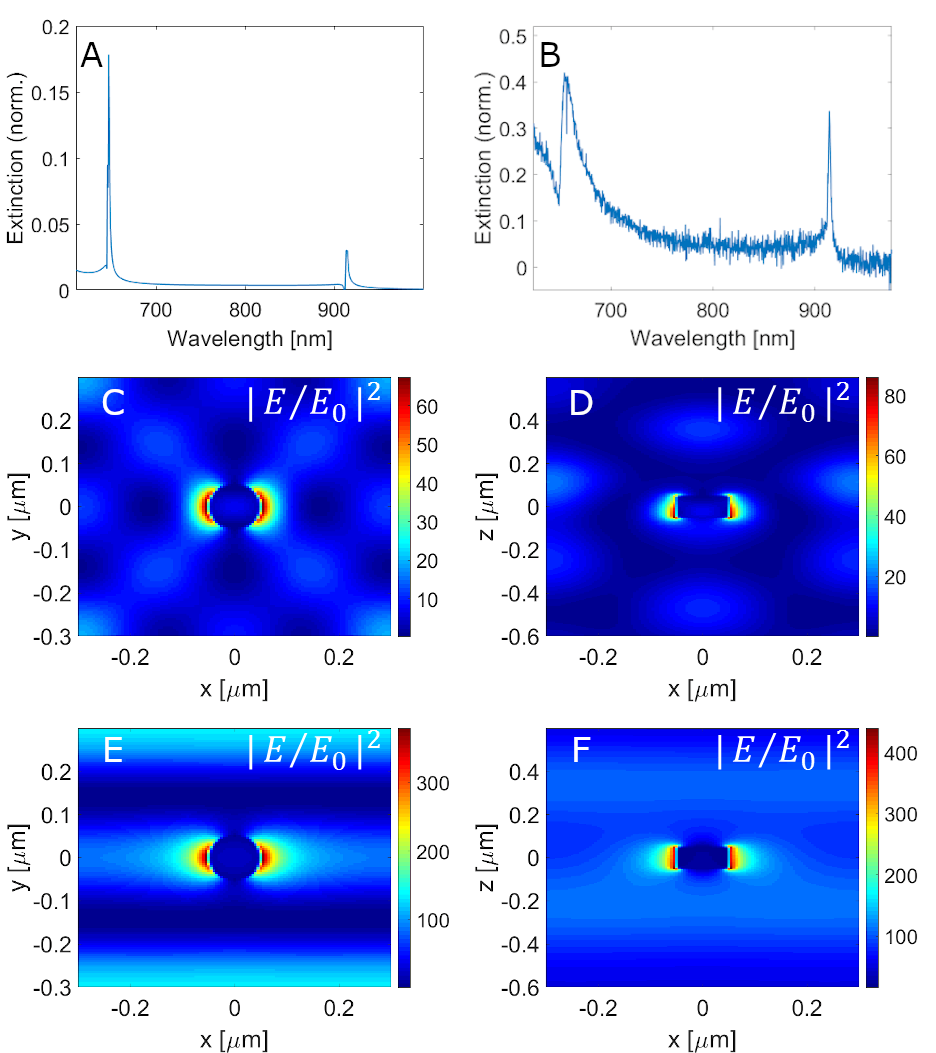}
\caption{FDTD simulation results for amorphous silicon. (A) Simulated transmission spectrum and (B) measured spectra at $k_{y}$=0 for comparison. Electric field intensity distribution at resonance at 646 nm crosscuts in the xy-plane (C) and xz-plane (D). Electric field intensity distribution at resonance at 913 nm crosscuts in the xy-plane (E) and xz-plane (F).}
\label{fig:sim_aSi}
\end{figure}
\begin{figure}[htbp!]
\centering
\includegraphics[width=0.8\textwidth]{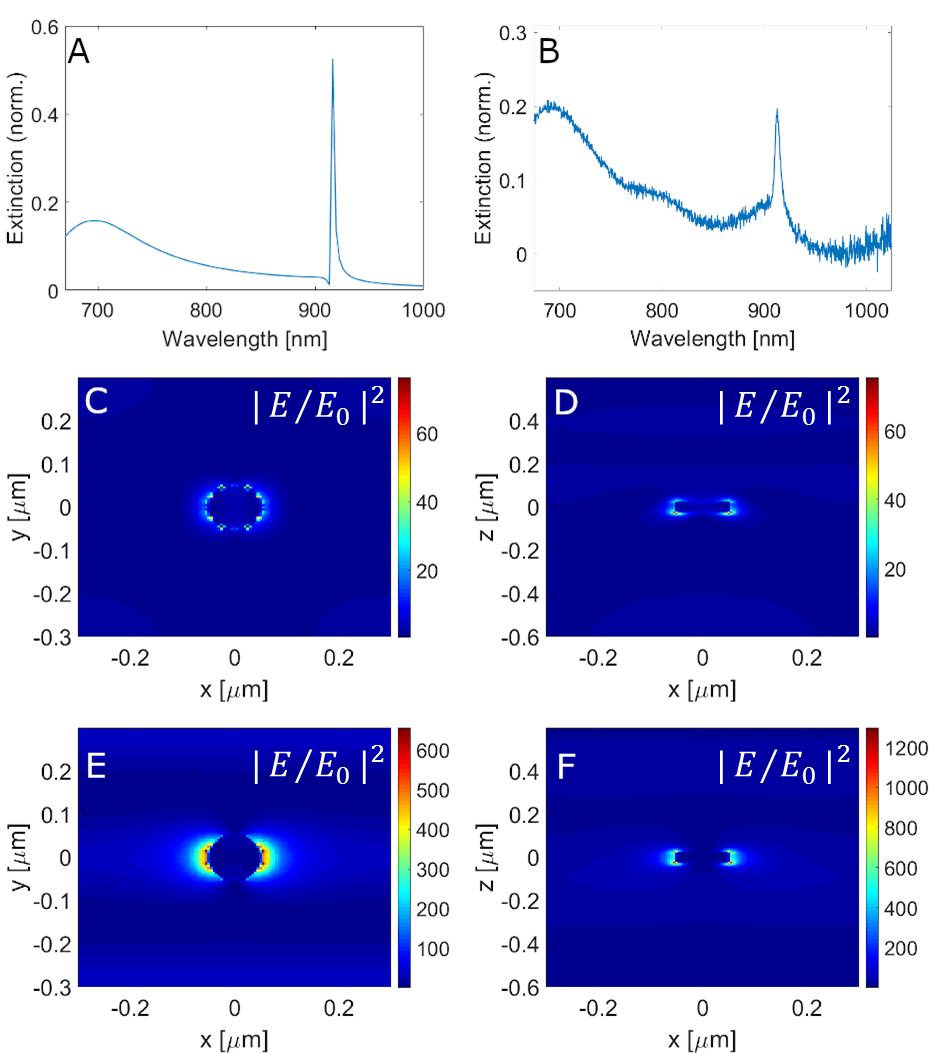}
\caption{FDTD simulation results for gold. (A) Simulated transmission spectrum and (B) measured spectra at $k_{y}$=0 for comparison. Electric field intensity distribution at resonance at 690 nm crosscuts in the xy-plane (C) and xz-plane (D). Electric field intensity distribution at resonance at 916 nm crosscuts in the xy-plane (E) and xz-plane (F). As it is difficult to see from the color scale, we note that the field does not decay to negligible values within the shown $z$-range.}
\label{fig:sim_Au}
\end{figure}

To investigate the near field distribution of the arrays and their possible connection to strong coupling, the structures are simulated using the FDTD-approach. The results are shown in Figure \ref{fig:sim_aSi} and Figure \ref{fig:sim_Au} for amorphous silicon and gold, respectively. For both materials, the electric field distribution is shown at the first and second SLR energies, which correspond to the first and second diffracted order, as crosscuts through the center of the nanoparticles in xy- and yz-planes. The first SLR is at 916 nm for gold and at 913 nm for amorphous silicon, while the second SLR is at 646 nm for gold and at 647 nm for amorphous silicon, see Figures \ref{fig:sim_aSi} A and \ref{fig:sim_Au} A. In the case of gold, the second SLR overlaps with the single particle resonance and therefore appears as a dip in the spectrum. The single particle resonances are located at 690 nm and 570 nm for gold and amorphous silicon, respectively. Figures \ref{fig:sim_aSi} B and \ref{fig:sim_Au} B show the measured extinction spectra for comparison to the simulations. In the case of amorphous silicon, the measured SLRs are located at 914 nm and 654 nm. For gold, the first SLR is located at 913 nm. The second SLR overlaps with the single particle resonance, which is at 691 nm. In general, the simulations and measurements agree very well. \par 
The local electric field intensity for the amorphous silicon is enhanced by a factor of 400 for the first SLR and by a factor of 80 for the second SLR. In the case of gold, the local electric field is enhanced by a factor of 1200 for the first SLR and by a factor of 70 close to the second SLR at 690 nm. The field enhancements are therefore larger for gold than for amorphous silicon for this set of parameters. Note that it is difficult to compare the field enhancement systematically since particles which are of the  same size, but made of different materials will have their single particle resonances at different energies. Nonetheless, the difference in the near field enhancement is probably the reason for the larger values for the Rabi splitting in case of the gold nanoparticle arrays.\par
The crosscuts through the electric field magnitude in the xy-plane (see Figures \ref{fig:sim_aSi} C, E and \ref{fig:sim_Au} C, E) and yz-plane (see Figures \ref{fig:sim_aSi} D, F and \ref{fig:sim_Au} D, F) are shown in a 1.2 $\mu$m range around the particle. In the case of gold nanoparticles, the field enhancement concentrates on the edges of the particle, while the case of amorphous silicon, the electric field has a substantial component inside the particle. Figures \ref{fig:sim_aSi} D, F and \ref{fig:sim_Au} D, F show that the fields also extend well over half a micron, despite field enhancement around the particles. This finding, supports the observation that strong coupling of the SLRs to the dye molecules is possible in thicker films even at 25 mM dye concentration. The simulations were run without the titanium disk beneath the gold nanoparticles, as we found that the titanium does not significantly affect the results (not shown).\par 

\section{Discussion}
In summary, we have studied light-matter interaction between all-dielectric nanoparticle arrays consisting of amorphous silicon and organic dye molecules, and compared them to a similar system of gold nanoparticle arrays. We found that the SLRs of the dielectric nanoarrays form strong coupling with the emitters, which has, to our knowledge, not been observed before. We show a linear dependence of the Rabi splitting on the dye molecule concentration.  We found Rabi splittings up to 253 eV and 293 eV for amorphous silicon and gold, respectively. In our work we used square arrays of cylinders. Recently Todisco {\it et al.}\cite{todisco_toward_2016}
demonstrated that further increases in Rabi splittings can be achieved by using
disk dimers with narrow gap in between. This is due to stronger field localization
between disks when incident light polarization is along the dimer. It would be interesting
to explore similar effects in our systems.
We also found that the Rabi splitting which was not observed at a concentration of 25 mM and a film thickness of 150 nm can be observed at film thicknesses of around 540 nm. The simulated near-field enhancements were up to a factor of 1200 close to the particles in the case of gold nanoparticles and 400 for amorphous silicon nanoparticles. The simulations show that the modes extend considerably beyond the particles in the substrate normal direction. \par
The simulations and measurements generally agree very well. The SLRs in the measurements are red-shifted by approximately 3 nm with respect to the SLRs in the FDTD simulations. This difference is most likely due to small irregularities in the fabricated arrays and a change in the refractive index introduced by the presence of the dye molecules and the PMMA film. \par
In our system, the resonances of amorphous silicon nanoarrays show similar behaviour in the strong coupling regime and provide similar linewidths and quality factors for the resonant modes as the gold nanoarrays. However, it has been proposed that dielectric nanostructures could provide much narrower linewidths due to their non-dissipative nature ~\cite{krasnok2018spectroscopy}. The linewidths are strongly dependent on size and shape of the particles ~\cite{schuller2009general} as well as on the location of the SLR mode relative to the single particle resonance \cite{nikitin2012plasmonic,li_engineering_2018}. The linewidth is strongly related to the lifetime of the  mode and plays therefore an important role in coherence phenomena such as lasing. Strongly coupled systems of dielectric nanoarrays can therefore lead to new developments and improvements in lasing applications, for example toward polariton lasers and condensates with lower thresholds.

\section{Acknowledgements}
\textbf{Funding:}
This  work  was
supported by the Academy of Finland under project numbers 303351,  307419, 327293, 318987 (QuantERA project RouTe), and 318937 (PROFI), by Centre for Quantum Engineering (CQE) at Aalto University,
and by the European Research Council (ERC-2013-AdG-340748-CODE).  Part of the research was performed at the
Micronova Nanofabrication Centre, supported by Aalto University.
\textbf{Competing interests:} The authors declare no competing financial interests. \textbf{Data and materials availability:} All data needed to evaluate the conclusions in the paper are present in the paper.

\bibliography{dielectricspaper}

\begin{thebibliography}{10}

\bibitem{torma_strong_2015}
P~Törmä and W~L Barnes.
\newblock Strong coupling between surface plasmon polaritons and emitters: a
  review.
\newblock {\em Reports on Progress in Physics}, 78(1):013901, January 2015.

\bibitem{kravets2018plasmonic}
Vasyl~G Kravets, Andrei~V Kabashin, William~L Barnes, and Alexander~N
  Grigorenko.
\newblock Plasmonic surface lattice resonances: a review of properties and
  applications.
\newblock {\em Chemical reviews}, 118(12):5912--5951, 2018.

\bibitem{guebrou2012coherent}
S~Aberra Guebrou, C~Symonds, E.~Homeyer, et~al.
\newblock Coherent emission from a disordered organic semiconductor induced by
  strong coupling with surface plasmons.
\newblock {\em Physical Review Letters}, 108(6):066401, 2012.

\bibitem{shi_spatial_2014}
L.~Shi, T.~K. Hakala, H.~T. Rekola, J.-P. Martikainen, R.~J. Moerland, and
  P.~Törmä.
\newblock Spatial coherence properties of organic molecules coupled to
  plasmonic surface lattice resonances in the weak and strong coupling regimes.
\newblock {\em Physical Review Letters}, 112(15), April 2014.
\newblock arXiv: 1404.4160.

\bibitem{vakevainen2019sub}
Aaro~I V{\"a}kev{\"a}inen, Antti~J Moilanen, Marek Ne{\v{c}}ada, Tommi~K
  Hakala, Konstantinos~S Daskalakis, and P{\"a}ivi T{\"o}rm{\"a}.
\newblock Sub-picosecond thermalization dynamics in condensation of strongly
  coupled lattice plasmons.
\newblock {\em arXiv preprint arXiv:1905.07609}, 2019.

\bibitem{orgiu2015conductivity}
E~Orgiu, J~George, JA~Hutchison, et~al.
\newblock Conductivity in organic semiconductors hybridized with the vacuum
  field.
\newblock {\em Nature Materials}, 14(11):1123, 2015.

\bibitem{chervy2016high}
Thibault Chervy, Jialiang Xu, Yulong Duan, et~al.
\newblock High-efficiency second-harmonic generation from hybrid light-matter
  states.
\newblock {\em Nano letters}, 16(12):7352--7356, 2016.

\bibitem{ribeiro2018polariton}
Raphael~F Ribeiro, Luis~A Mart{\'\i}nez-Mart{\'\i}nez, Matthew Du, Jorge
  Campos-Gonzalez-Angulo, and Joel Yuen-Zhou.
\newblock Polariton chemistry: controlling molecular dynamics with optical
  cavities.
\newblock {\em Chemical science}, 9(30):6325--6339, 2018.

\bibitem{hertzog2019strong}
Manuel Hertzog, Mao Wang, J{\"u}rgen Mony, and Karl B{\"o}rjesson.
\newblock Strong light--matter interactions: a new direction within chemistry.
\newblock {\em Chemical Society Reviews}, 48(3):937--961, 2019.

\bibitem{de2007colloquium}
FJ~Garcia De~Abajo.
\newblock Colloquium: Light scattering by particle and hole arrays.
\newblock {\em Reviews of Modern Physics}, 79(4):1267, 2007.

\bibitem{bozhevolnyi2017quantum}
Robert~J. Moerland, Tommi~K. Hakala, Jani-Petri Martikainen, Heikki~T. Rekola,
  Aaro~I. V{\"a}kev{\"a}inen, and P{\"a}ivi T{\"o}rm{\"a}.
\newblock Strong coupling between organic molecules and plasmonic
  nanostructures.
\newblock In Sergey~I. Bozhevolnyi, Luis Martin-Moreno, and Francisco
  Garcia-Vidal, editors, {\em Quantum Plasmonics}, pages 121--150, Cham, 2017.
  Springer International Publishing.

\bibitem{wang2018rich}
Weijia Wang, Mohammad Ramezani, Aaro~I V{\"a}kev{\"a}inen, P{\"a}ivi
  T{\"o}rm{\"a}, Jaime~G{\'o}mez Rivas, and Teri~W Odom.
\newblock The rich photonic world of plasmonic nanoparticle arrays.
\newblock {\em Materials today}, 21(3):303--314, 2018.

\bibitem{vakevainen_plasmonic_2014}
A.~I. Väkeväinen, R.~J. Moerland, H.~T. Rekola, et~al.
\newblock Plasmonic surface lattice resonances at the strong coupling regime.
\newblock {\em Nano Letters}, 14(4):1721--1727, April 2014.

\bibitem{ramezani2017plasmon}
Mohammad Ramezani, Alexei Halpin, Antonio~I Fern{\'a}ndez-Dom{\'\i}nguez,
  et~al.
\newblock Plasmon-exciton-polariton lasing.
\newblock {\em Optica}, 4(1):31--37, 2017.

\bibitem{de2018interaction}
Milena De~Giorgi, Mohammad Ramezani, Francesco Todisco, et~al.
\newblock Interaction and coherence of a plasmon--exciton polariton condensate.
\newblock {\em Acs Photonics}, 5(9):3666--3672, 2018.

\bibitem{staude2017metamaterial}
Isabelle Staude and J{\"o}rg Schilling.
\newblock Metamaterial-inspired silicon nanophotonics.
\newblock {\em Nature Photonics}, 11(5):274, 2017.

\bibitem{kuznetsov_optically_2016}
Arseniy~I. Kuznetsov, Andrey~E. Miroshnichenko, Mark~L. Brongersma, Yuri~S.
  Kivshar, and Boris Luk’yanchuk.
\newblock Optically resonant dielectric nanostructures.
\newblock {\em Science}, 354(6314):aag2472, November 2016.

\bibitem{evlyukhin2010optical}
Andrey~B Evlyukhin, Carsten Reinhardt, Andreas Seidel, Boris~S Luk’yanchuk,
  and Boris~N Chichkov.
\newblock Optical response features of si-nanoparticle arrays.
\newblock {\em Physical Review B}, 82(4):045404, 2010.

\bibitem{krasnok2018spectroscopy}
Alex Krasnok, Mart{\'\i}n Caldarola, Nicolas Bonod, and Andrea Al{\'u}.
\newblock Spectroscopy and biosensing with optically resonant dielectric
  nanostructures.
\newblock {\em Advanced optical materials}, 6(5), 2018.

\bibitem{tserkezis2018mie}
Christos Tserkezis, PAD Gon{\c{c}}alves, Christian Wolff, Francesco Todisco,
  Kurt Busch, and N~Asger Mortensen.
\newblock Mie excitons: Understanding strong coupling in dielectric
  nanoparticles.
\newblock {\em Physical Review B}, 98(15):155439, 2018.

\bibitem{wang2016resonance}
Hao Wang, Yanlin Ke, Ningsheng Xu, et~al.
\newblock Resonance coupling in silicon nanosphere--j-aggregate
  heterostructures.
\newblock {\em Nano letters}, 16(11):6886--6895, 2016.

\bibitem{liu2018resonance}
Shao-Ding Liu, Jin-Li Fan, Wen-Jie Wang, Jing-Dong Chen, and Zhi-Hui Chen.
\newblock Resonance coupling between molecular excitons and nonradiating
  anapole modes in silicon nanodisk-j-aggregate heterostructures.
\newblock {\em ACS Photonics}, 5(4):1628--1639, 2018.

\bibitem{todisco2019magnetic}
Francesco Todisco, Radu Malureanu, Christian Wolff, et~al.
\newblock Magnetic and electric mie-exciton polaritons in silicon nanodisks.
\newblock {\em arXiv preprint arXiv:1906.09898}, 2019.

\bibitem{lumerical}
Lumerical {Inc}.
\newblock \url{https://www.lumerical.com/products/}.

\bibitem{lide1995crc}
David~R Lide.
\newblock {\em CRC handbook of chemistry and physics: a ready-reference book of
  chemical and physical data}.
\newblock CRC press, 1995.

\bibitem{li_engineering_2018}
Jiaqi Li, Niels Verellen, and Pol Van~Dorpe.
\newblock Engineering electric and magnetic dipole coupling in arrays of
  dielectric nanoparticles.
\newblock {\em Journal of Applied Physics}, 123(8):083101, February 2018.

\bibitem{ruppin2002electromagnetic}
R~Ruppin.
\newblock Electromagnetic energy density in a dispersive and absorptive
  material.
\newblock {\em Physics letters A}, 299(2-3):309--312, 2002.

\bibitem{koenderink2010use}
A~Femius Koenderink.
\newblock On the use of purcell factors for plasmon antennas.
\newblock {\em Optics letters}, 35(24):4208--4210, 2010.

\bibitem{gonzalez2013theory}
A~Gonz{\'a}lez-Tudela, PA~Huidobro, Luis Mart{\'\i}n-Moreno, C~Tejedor, and
  FJ~Garc{\'\i}a-Vidal.
\newblock Theory of strong coupling between quantum emitters and propagating
  surface plasmons.
\newblock {\em Physical review letters}, 110(12):126801, 2013.

\bibitem{winkler2018room}
Jan~M Winkler, Freddy~T Rabouw, Aurelio~A Rossinelli, et~al.
\newblock Room-temperature strong coupling of cdse nanoplatelets and plasmonic
  hole arrays.
\newblock {\em Nano letters}, 19(1):108--115, 2018.

\bibitem{todisco_toward_2016}
Francesco Todisco, Marco Esposito, Simone Panaro, Milena De~Giorgi, Lorenzo
  Dominici, Dario Ballarini, Antonio~I. Fern{\'a}ndez-Dom{\'i}nguez,
  Vittorianna Tasco, Massimo Cuscun{\`a}, Adriana Passaseo, Cristian
  Cirac{\`i}, Giuseppe Gigli, and Daniele Sanvitto.
\newblock Toward {Cavity} {Quantum} {Electrodynamics} with {Hybrid} {Photon}
  {Gap}-{Plasmon} {States}.
\newblock {\em ACS Nano}, 10(12):11360--11368, 2016.

\bibitem{schuller2009general}
Jon~A Schuller and Mark~L Brongersma.
\newblock General properties of dielectric optical antennas.
\newblock {\em Optics express}, 17(26):24084--24095, 2009.

\bibitem{nikitin2012plasmonic}
Andrey~G Nikitin, Andrei~V Kabashin, and Herv{\'e} Dallaporta.
\newblock Plasmonic resonances in diffractive arrays of gold nanoantennas: near
  and far field effects.
\newblock {\em Optics express}, 20(25):27941--27952, 2012.

\end{thebibliography}

\newpage
\begin{table}[htbp!]
\centering
\caption{Measured linewidth of the dye molecules, $\gamma_{\mathrm{dye}}$, the strong coupling condition, $(\gamma_{\mathrm{dye}} + \gamma_{\mathrm{SLR}})/2$, and Rabi splittings of amorphous silicon and gold, $\Omega_{\mathrm{a-Si}}$ and $\Omega_{\mathrm{Au}}$, for different concentrations, $c$, for a film thickness of 150 nm. The molecule linewidth is obtained by fitting two Gaussians to the measured absorption spectrum (see Figure \ref{fig:setup}), and choosing the underlying Gaussian for the main absorption peak.}
\label{tab:splitting}
\begin{tabular}{ |c|c|c|c|c| }
 \hline
 c [mM] & $\gamma_{\mathrm{dye}}$ [eV] &$(\gamma_{\mathrm{dye}} + \gamma_{\mathrm{SLR}})/2$ [eV] &$\Omega_{\mathrm{a-Si}}$ [eV] & $\Omega_{\mathrm{Au}}$ [eV] \\ \hline
 25 & 0.10 & 0.06 & - & - \\ 
 50 & 0.11 & 0.06 & 0.13 & 0.13\\ 
 75 & 0.12 & 0.07 & 0.16 & 0.19\\ 
 100 & 0.11 & 0.06 & 0.17 & 0.20\\ 
 150 & 0.12 & 0.07 & 0.21 & 0.23\\ 
 200 & 0.13 & 0.07 & 0.25 & 0.29\\ 
 \hline
\end{tabular}
\end{table}
\newpage
\newpage
\newpage
\newpage
\newpage
\clearpage

\section*{Supplementary Material}
\setcounter{figure}{0} 
\setcounter{table}{0} 

\begin{table}[htbp!]
	\centering
	\caption{Measured linewidth of the dye molecules, $\gamma_{\mathrm{dye}}$, the strong coupling condition, $(\gamma_{\mathrm{dye}} + \gamma_{\mathrm{SLR}})/2$, and Rabi splittings of amorphous silicon and gold, $\Omega_{\mathrm{a-Si}}$ and $\Omega_{\mathrm{Au}}$, for different concentrations, $c$, for a film thickness of 150 nm. The molecule linewidth is obtained by fitting one Gaussian to the measured absorption spectrum.}
	\label{tab:splitting_SI}
	\begin{tabular}{ |c|c|c|c|c| }
		\hline
		c [mM] & $\gamma_{\mathrm{dye}}$ [eV] &$(\gamma_{\mathrm{dye}} + \gamma_{\mathrm{SLR}})/2$ [eV] &$\Omega_{\mathrm{a-Si}}$ [eV] & $\Omega_{\mathrm{Au}}$ [eV] \\ \hline
		25 & 0.14 & 0.08 & - & - \\ 
		50 & 0.17 & 0.09 & 0.18 & 0.19\\ 
		75 & 0.20 & 0.11 & 0.24 & 0.25\\ 
		100 & 0.20 & 0.11 & 0.24 & 0.27\\ 
		150 & 0.23 & 0.12 & 0.30 & 0.31\\ 
		200 & 0.28 & 0.15 & 0.36 & 0.39\\ 
		\hline
	\end{tabular}
\end{table}

\begin{figure}[htbp!]
	\centering
	\includegraphics[width=0.8\textwidth]{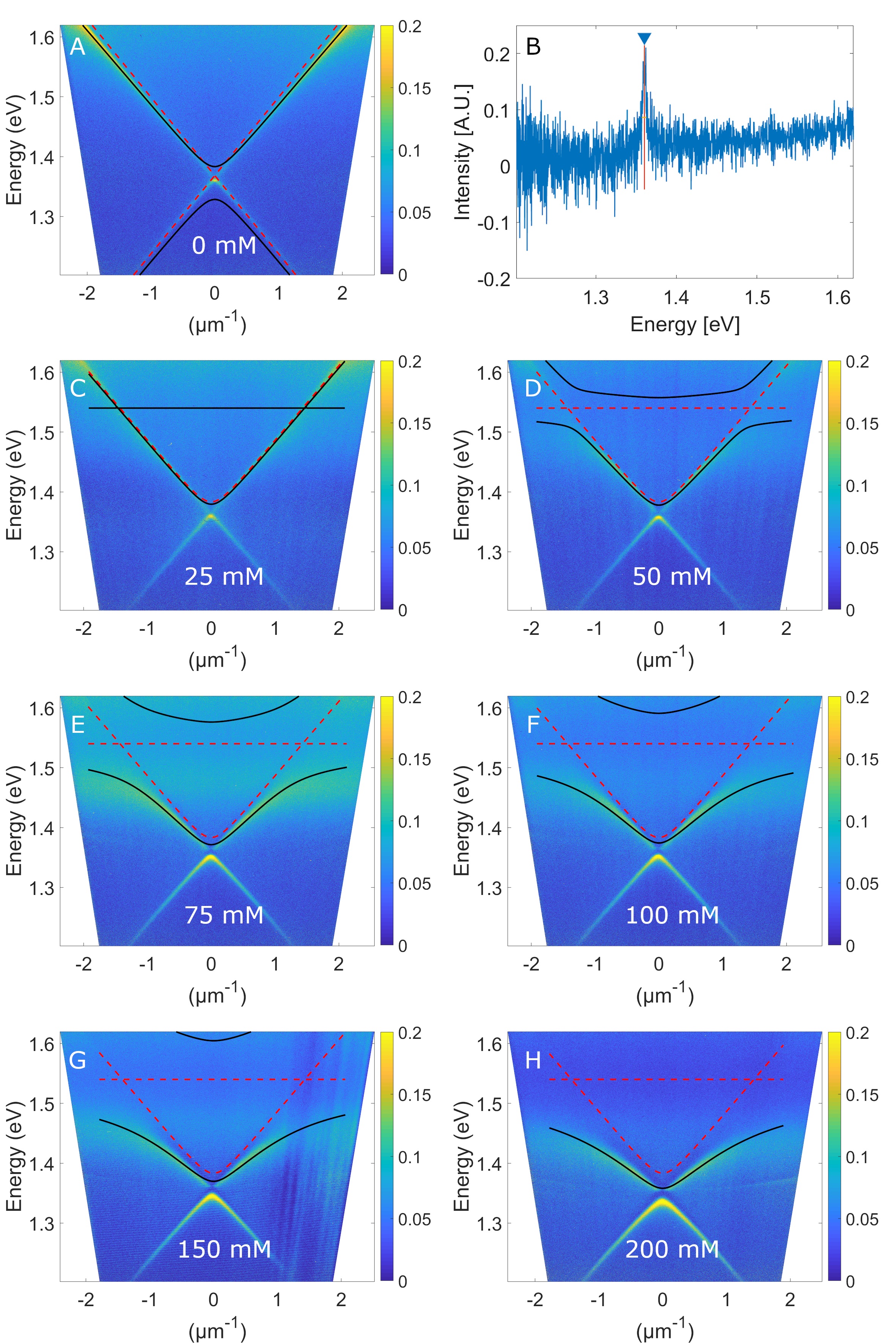}
	\caption{Dispersion measurements of a nanoarray of gold nanoparticles with a periodicity of 600 nm in x- and y-direction, diameter of 100 nm and dye concentrations of (A) 0 mM (C) 25 mM, (D) 50 mM, (E) 75 mM, (F) 100 mM, (G) 150 mM and (H) 200 mM. The film thickness is 150 nm. Maximum extinction is given in yellow. (B) shows a crosscut through (A) at $k_{y}=0$. In (A) the red dashed lines depict the uncoupled diffracted orders of the lattice and the black solid lines the SLRs obtained by the coupled mode fitting of the extinction maxima. In (C), (D), (E), (F), (G) and (H) the horizontal red dashed line depicts the absorption line of the dye molecules and the diagonal red lines correspond to the fitted SLRs (upper branch) from Figure \ref{fig:splitting} (A). The solid black lines depict the coupled mode fitting of the extinction maxima. As can be seen the Rabi splitting increases with increasing dye molecule concentration.}
	\label{fig:splitting_Au}
\end{figure}
\newpage
\begin{figure}[htbp!]
	\centering
	\includegraphics[width=1\textwidth]{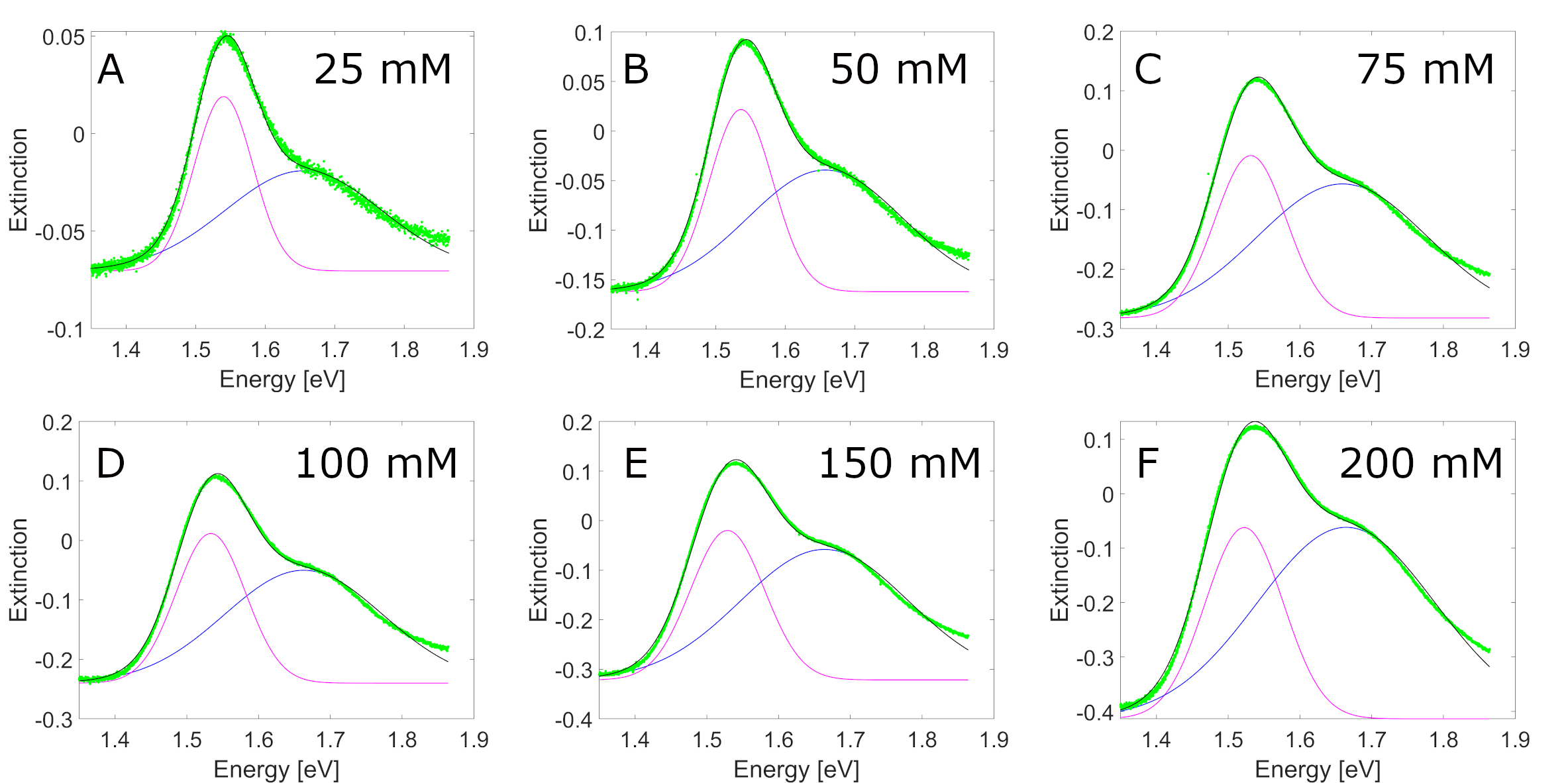}
	\caption{Absorption spectra of IR-792 solved in PMMA A3 with dye concentrations of (A) 25 mM, (B) 50 mM, (C) 75 mM, (D) 100 mM, (E) 150 mM and (F) 200 mM. We fit two Gaussians to the spectrum, shown by the pink and blue curves. The pink curve corresponds to the main absorption peak.}
	\label{fig:absorption}
\end{figure}

\begin{figure}[htbp!]
	\centering
	\includegraphics[width=1\textwidth]{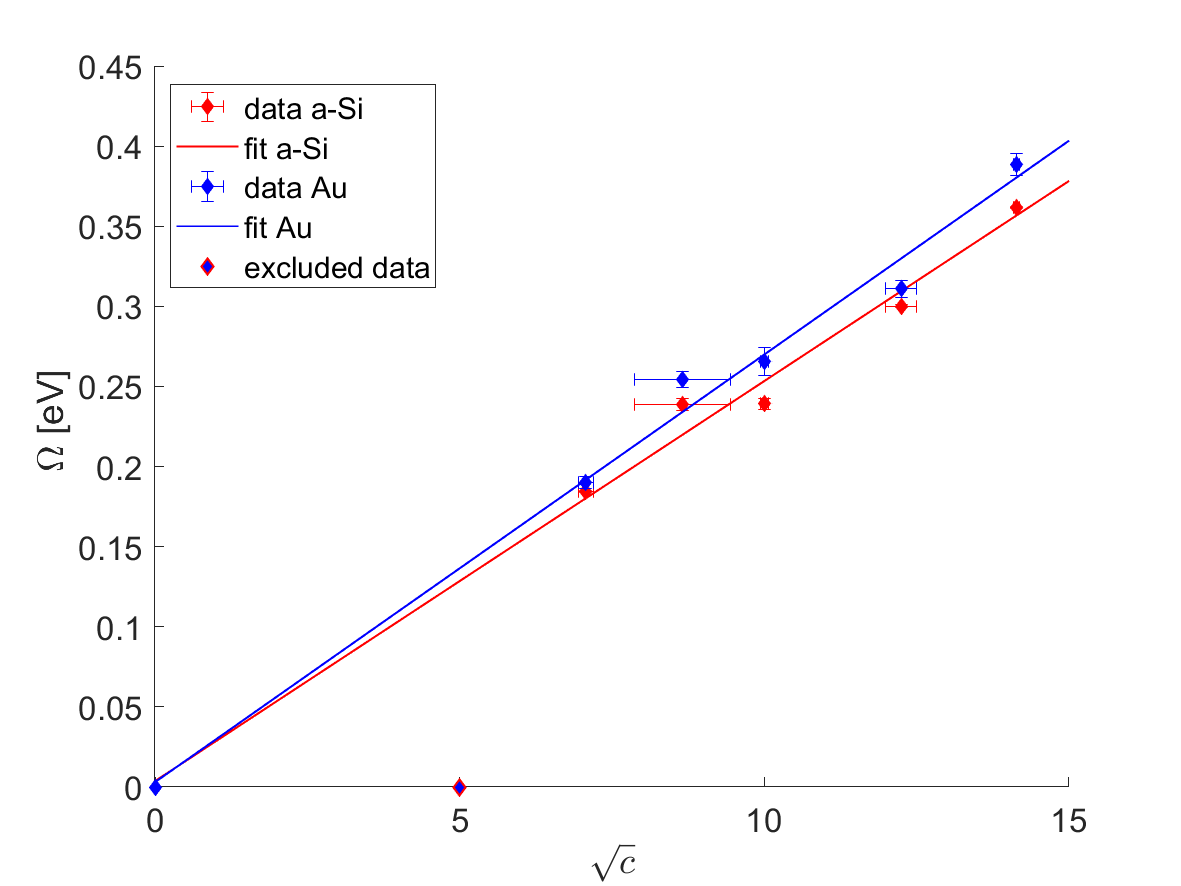}
	\caption{Dependence of the Rabi splitting on the square root of the concentration $\sqrt{c}$ for a film thickness of 150 nm. The data point $c=25$ mM is excluded since the splitting is smaller than the linewidth and thus not measurable. For this fit, the molecule linewidth is obtained by fitting one Gaussian to the measured absorption spectrum. For corresponding values, see Table \ref{tab:splitting_SI}.}
	\label{fig:splitting_1gaussian_SI}
\end{figure}

\end{document}